\documentclass[prb,twocolumn]{revtex4}
\usepackage{graphicx}

\begin{document}
\title{Anti-ferromagnetic ordering in arrays of superconducting $\pi$-rings}
\author{
J.R. Kirtley$^1$,
C.C. Tsuei$^1$,
Ariando$^2$,
H.-J.H. Smilde$^2$,
H. Hilgenkamp$^2$,
}

\affiliation{$^{1}$IBM Watson Research Center, Yorktown Heights, NY, USA}
\affiliation{$^{2}$Low Temperature Division, Faculty of Science and Technology
and MESA$^+$ Research Institute, University of Twente, Enschede, The Netherlands}

\date{\today}

\begin{abstract}
We report experiments in which one dimensional (1D)
and two dimensional (2D) arrays of  YBa$_2$Cu$_3$O$_{7-\delta}$-Nb $\pi$-rings are cooled
through the superconducting transition temperature of the Nb in various magnetic fields.
These $\pi$-rings have
degenerate ground states with either clockwise or counter-clockwise
spontaneous circulating supercurrents.
The final flux state of each ring in the
arrays was determined using scanning SQUID microscopy. In the 1D arrays, fabricated
as a single junction with facets alternating between alignment parallel to a [100]
axis of the YBCO and rotated $90^o$ to that axis,
half-fluxon Josephson vortices
order strongly into an arrangement with alternating signs of their magnetic flux.
We demonstrate that this ordering is driven by phase coupling
and model the cooling process with a numerical solution of the Sine-Gordon equation.
The 2D ring arrays couple to each other
through the magnetic flux generated by the spontaneous supercurrents. Using $\pi$-rings for
the 2D flux coupling experiments eliminates one source of disorder seen in similar
experiments using conventional superconducting rings,
since $\pi$-rings have doubly degenerate ground states in
the absence of an applied field. Although anti-ferromagnetic ordering occurs, with larger negative
bond orders than previously reported for arrays of conventional rings,
long-range order is never observed, even in geometries without geometric frustration.
This may be due to dynamical effects. Monte-Carlo simulations of the
2D array cooling process are presented and compared with experiment.
\end{abstract}

\maketitle



\section{Introduction}

In superconducting rings,
the requirement of a single-valued, macroscopic quantum mechanical wave-function,
combined with
the intimate relation between the quantum mechanical phase and the
vector potential, results in flux quantization.\cite{deaver1961,doll1961}
Under the appropriate conditions, two of the flux quantized states can
become degenerate,
characterized by time-reversed, persistent supercurrents circulating in a clockwise
and counter-clockwise direction. These macroscopic circulating currents can be
thought of as an analog to an electronic spin.
In pioneering work, Davidovic et al.\cite{davidovic1996, davidovic1997} showed
that arrays of superconducting rings can be used as models for spin systems,
since they interact anti-ferromagnetically upon cooling. However, the Davidovic
arrays never showed long-range anti-ferromagnetic ordering. They speculated
that one reason for this lack of ordering is that, for there to be two degenerate
states, such rings must be cooled in a field
equivalent to a half-integer multiple of the superconducting flux quantum $\Phi_0=h/2e$ per ring.
Therefore variations in the lithographically patterned areas of these rings result
in different fluxes through different rings in the same field, causing disorder.
Superconducting $\pi$-rings have an intrinsic phase shift of $\pi$ in the
absence of an externally applied field or supercurrent.
\cite{bulaevskii1977,geshkenbein1986,geshkenbein1987} Such rings
have a doubly degenerate ground state in zero applied flux, should
not have this source of disorder, and may therefore be a more ideal model for
a spin system.

However, use of $\pi$-rings for a model spin system requires a large number of
rings. The first superconducting  $\pi$-rings,\cite{wollman1993,brawner1994,tsuei1994,mathai1994}
which depended on the momentum
dependence of the pairing wavefunction in the high-T$_c$ cuprate perovskite superconductors, were
made using technologies which would be difficult to extend to many devices.
Recently a technology that allows for photolithographic fabrication
of  $\pi$-shift devices and arrays of great complexity, using YBa$_2$Cu$_3$O$_{7-\delta}$-Au-Nb
(YBCO-Au-Nb) ramp-edge tunneling junctions has been demonstrated.\cite{smilde2002a,smilde2002b}
Moreover, it has recently been demonstrated that  $\pi$-rings can also be fabricated
using Josephson junctions with ferromagnetic layers in the tunnel barriers.
\cite{andreev1991,ryazanov2002,bauer2004}
In this paper we report on experiments in which the YBCO ramp-edge technology was used to
fabricate large arrays of $\pi$-rings. A first report on work with similar arrays appeared
in Ref. \onlinecite{hilgenkamp2003}. The arrays were cooled in various magnetic fields,
and the final ``spin" states of the arrays were determined with  scanning SQUID microscopy.
Long range anti-ferromagnetic (AFM) ordering was observed in the 1D arrays. Although stronger
anti-ferromagnetic correlations were observed in our 2D $\pi$-ring arrays than were reported
previously for the conventional ($0$-ring) arrays, in neither case did
AFM ordering extend beyond a few lattice sites. Although using $\pi$-rings
eliminated one source of disorder, inhomogeneous flux biasing due to the applied fields
required for $0$-rings,
there are other sources of disorder, including inhomogeneities in the ring
critical currents and critical temperatures. We will discuss the influences of these
sources on our results, as well as dynamic effects, using Monte-Carlo modelling of the
cooling process.

\begin{figure}
\includegraphics[width=3.5in]{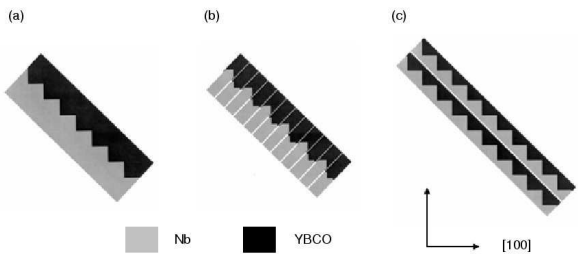}
\vspace{0.1in}
\caption{Schematics of the 3 types of facetted junctions studied in this work.
(a) A continuous, single facetted junction. (b) Junction with the individual half-fluxons
electrically isolated from one another. (c) Double row of continuous facetted junctions.}
\label{fig:hilgfacets}
\end{figure}

\section{Experimental methods}

Various samples consisting of one- and two-dimensional $\pi$-ring
arrays have been realized using ramp-type
YBa$_2$Cu$_3$O$_{7-\delta}$ - Au - Nb Josephson contacts. The
fabrication of ramp-type YBa$_2$Cu$_3$O$_{7-\delta}$ - Au - Nb
Josephson has been described previously in detail in Ref.'s
\onlinecite{smilde2002a,smilde2002b}. In short, the samples were
prepared by first epitaxially growing a bilayer of [001]-oriented
YBa$_2$Cu$_3$O$_{7-\delta}$ and SrTiO$_3$ by pulsed-laser
deposition on [001]-oriented SrTiO$_3$ single crystal substrates.
For the 1D array samples a 150~nm YBa$_2$Cu$_3$O$_{7-\delta}$
and a 100~nm SrTiO$_3$ film were used, while for the 2D array samples
a 340~nm YBa$_2$Cu$_3$O$_{7-\delta}$ and a 67~nm SrTiO$_3$ film were used.
In these bilayers the basic layout of the structures, which will
be described below in more detail, is defined by photolithography
and Ar ion milling. This process results in interfaces with a
slope of $15-35^\circ$ with respect to the substrate, which
provides access to the $ab$-planes of the
YBa$_2$Cu$_3$O$_{7-\delta}$ and allows the exploitation of
$d$-wave phase effects. Special care is taken to align all
interfaces accurately along one of the YBa$_2$Cu$_3$O$_{7-\delta}$
[100] axes. After etching the ramp and cleaning of the sample, a
7~nm YBa$_2$Cu$_3$O$_{7-\delta}$ interlayer is deposited, the
function and properties of which are described in Ref.
\onlinecite{smilde2002b}, followed by the in-situ pulsed-laser
deposition of a Au barrier-layer of 12~nm for the 1D array
samples and 20~nm for the 2D array samples. A 160 nm Nb counter
electrode is then formed by dc sputtering and structured by
lift-off. Each chip contained several reference-junctions. At T =
4.2 K, these showed a typical critical current per micrometer
junction-width $I_c/w\approx0.10$~mA/$\mu$m for the 1D array
samples and $I_c/w\approx0.18$~mA/$\mu$m for the 2D array
samples. From these, a value for the Josephson penetration depth
$\lambda_j \approx 1\mu$m (T = 4.2 K) for all samples is deduced,
which is the characteristic length scale over which Josephson
vortices (fluxons) extend.

\begin{figure}
\includegraphics[width=3.5in]{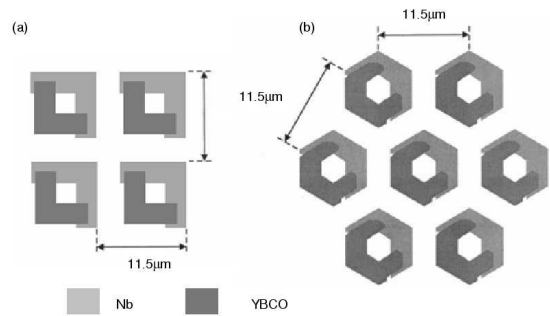}
\vspace{0.1in}
\caption{Schematic diagram of two types of $\pi$-rings used for the 2D arrays
reported in this paper. Type (a)
was used in the square arrays, and type (b) was used in the triangular, honeycomb,
and kagom{\'e} arrays.}
\label{fig:ec1schem}
\end{figure}

The sample magnetic fields were imaged with a
high resolution scanning SQUID microscope.\cite{rogers1983,vu1993,black1993,kirtley1995}
The SQUID microscope images
shown here were made at a temperature of T $<$ 5K, with the sample cooled
and imaged in the same fields. The size of the pickup loop used will be indicated for each image.
The samples were warmed through the superconducting critical temperature of the Nb and cooled
at controlled rates either using a non-inductive heater, or by passing warm $^4$He gas past
the sample.

\begin{figure}
\includegraphics[width=3.5in]{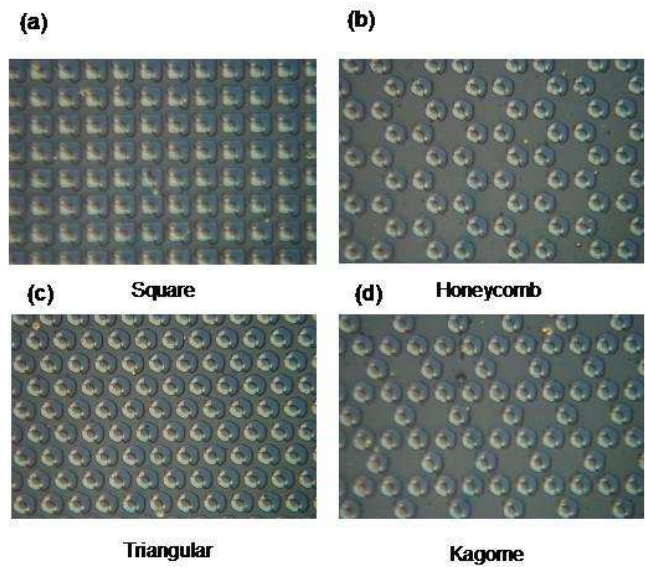}
\caption{Scanning electron microscopy images of 4 arrays of $\pi$-rings, with 2.7$\mu$m
junctions and 11.5$\mu$m ring to ring spacings.}
\label{fig:ec1sems}
\end{figure}

A first configuration for which the generation and coupling of half-integer
flux quanta was investigated is the zigzag array,\cite{smilde2002a} several instances of which
are shown schematically in Figure \ref{fig:hilgfacets}.
In these structures, the $d$-wave order parameter of the
YBa$_2$Cu$_3$O$_{7-\delta}$ induces a difference of $\pi$ in the Josephson phase-shift $\phi$  across the
YBa$_2$Cu$_3$O$_{7-\delta}$-Au-Nb barrier for neighboring facets. For facet lengths $a$ in the wide
limit, i.e., $a  >> \lambda_j$, the lowest-energy ground state of the system is expected
to be characterized by a spontaneous generation of a half-integer flux-quantum
at each corner. This half-fluxon provides a further  $\pi$-phase change between neighboring
facets, either adding or subtracting to the $d$-wave induced  $\pi$-phase shift, depending
on the half-flux quantum polarity. In both cases this leads to a lowering of the
Josephson coupling energy across the barrier, as this energy is proportional to (1 - cos$\phi$ ).
We studied three types of facetted junctions.
The first (Fig. \ref{fig:hilgfacets}a) was
an isolated, continuous junction with many adjacent facets. The second (Fig. \ref{fig:hilgfacets}b)
was lithographically patterned to electrically isolate each half-fluxon from its neighbor. The final type
(Fig. \ref{fig:hilgfacets}c) had two continuous facetted junctions close together, but electrically
isolated from one another,
to test for field coupling between 1D arrays of half-fluxons.

\begin{table}[b]
\caption{Details of 2D lattice samples; critical current density is J$_c \approx 5\times 10^{8} A/m^2$}
\label{tab:lattice_details}
\begin{tabular} {|c|c|c|c|c|}
\hline
                            &Square            &Triangular &Honeycomb  &Kagom{\'e}      \\ \hline
height YBCO (nm)            &340               &340        &340        &340             \\ \hline
height STO (nm)             & 67               & 67        & 67        & 67             \\ \hline
height Nb (nm)              &160               &160        &160        &160             \\ \hline
width JJ 1 ($\mu$m)$^1$     &2.75              &2.70       &2.70       &2.70            \\ \hline
width JJ 2 ($\mu$m)$^1$     &2.75              &3.75       &3.75       &3.75            \\ \hline
hole ($\mu$m)               &2.50              &$\sim$3.10 &$\sim$3.10 &$\sim$3.10      \\ \hline
self-inductance (pH)$^2$    &3.93              &$\sim$3.90 &$\sim$3.90 &$\sim$3.90      \\ \hline
self-inductance (pH)$^3$    &4.57              &-          &-          &-               \\ \hline
nearest neighbor            &11.5              &11.5       &11.5       &11.5            \\
distance ($\mu$m)           &                  &           &           &                \\ \hline
nearest neighbor            &0.025             &-          &-          &-               \\
mutual (pH)$^3$             &                  &           &           &                \\ \hline
\multicolumn{5}{l} {$^1$ designed value}                                                 \\
\multicolumn{5}{l} {$^2$ estimate with standard formulas}                                 \\
\multicolumn{5}{l} {$^3$ estimate using FastHenry, $\lambda_{YBCO}$=160nm and $\lambda_{Nb}$=40nm} \\

\end{tabular}
\end{table}

The 2-dimensional $\pi$-ring arrays were made up of individual rings patterned as indicated
schematically in Figure \ref{fig:ec1schem}. There were two types of rings, square
(Fig. \ref{fig:ec1schem}a) and hexagonal (Fig. \ref{fig:ec1schem}b). The rings were
patterned into arrays with 4 different geometries, as shown in the scanning electron
microscopy images of Figure \ref{fig:ec1sems}. In all cases the nearest-neighbor distances
in these arrays was 11.5$\mu$m center to center.
The details of the ring
geometries, critical currents, self-inductances, and mutual-inductances are given in
Table 1.

\begin{figure}
\includegraphics[width=3.5in]{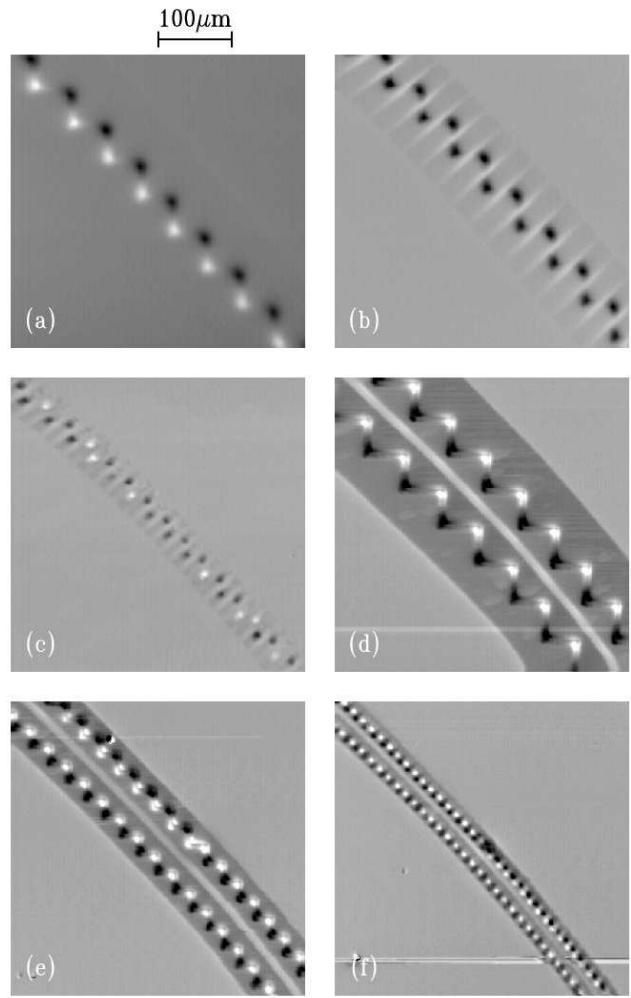}
\vspace{0.1in}
\caption{Scanning SQUID microscope images of zigzag YBCO-Au-Nb $0-\pi$
junctions, cooled in nominally zero field, and imaged with a 4$\mu$m diameter
pickup loop. (a) Continuous junction with 40$\mu$m facet lengths. (b)
Electrically disconnected junction with 40$\mu$m between facet corners. (c) Electrically
disconnected junction with 20$\mu$m between facet corners. Two parallel facetted $0-\pi$
junctions with 40$\mu$m (d), 20$\mu$m (e), and 10$\mu$m (f) between facet corners.
The apparent curvature of the junctions in these images is an artifact of
the scanning mechanism.}
\label{fig:hilgquad}
\end{figure}

\begin{figure}
\includegraphics[width=3.5in]{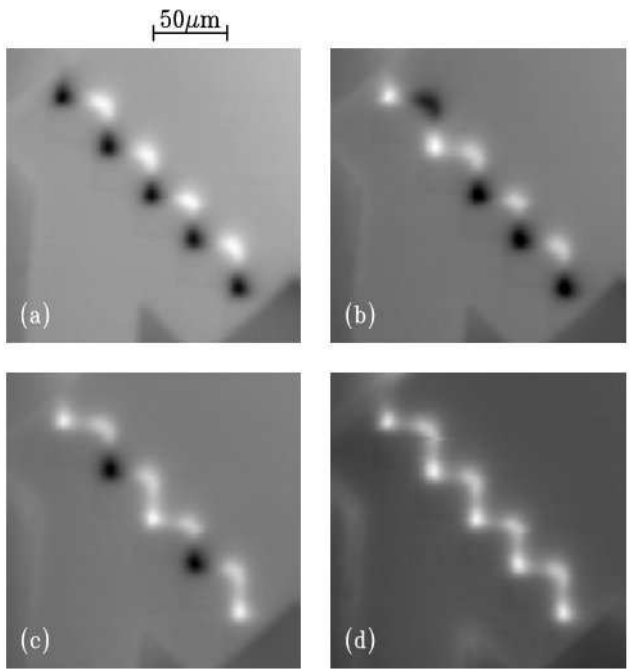}
\vspace{0.1in}
\caption{Scanning SQUID microscope images of a facetted YBCO-Nb $0-\pi$
junction with 10 facets each 40$\mu$m long, cooled in fields of
0nT (a), 32nT (b), 74nT (c), and 110nT (d), and imaged at 4.8K with an
8$\mu$m square pickup loop.}
\label{fig:hilgfacs}
\end{figure}

\section{Results}

Figure \ref{fig:hilgquad} shows representative scanning SQUID microscope images
of 6 zigzag $0-\pi$ facet junctions. All of these images were of samples
on the same
substrate, and imaged in the same cooldown in nominal zero field ( $< 0.5\mu$T)
at T = 4.2K. In all three types of facetted junctions, half-fluxon Josephson vortices
were spontaneously generated at the points where the facets met, as expected.
When the connected 1D
arrays (Fig. \ref{fig:hilgfacets}a) were cooled in zero field through the
niobium superconducting transition temperature,
the signs of the half-fluxons strongly tended towards perfect
anti-ferromagnetic ordering, in which the persistent supercurrents alternated
between clock-wise and counter clock-wise flow (Fig. \ref{fig:hilgquad}a,d,e,f).
There are two possible mechanisms for this ordering. In the first, ordering is
driven by the minimization of the total junction Josephson coupling energy
during the cooling process. We refer to this as ``phase" coupling, because the
Josephson currents and energies are determined by the difference in superconducting
phase across the junction $\phi$.
In the second, the anti-ferromagnetic ordering is driven by a minimization of the
total magnetic field energy in the junction and its environment.
We refer to this as ``field" coupling. In order to determine the
relative strengths of these mechanisms, facetted junctions were fabricated with
each half-fluxon electrically isolated from its neighbor (Fig. \ref{fig:hilgfacets}b).
This should eliminate the phase coupling mechanism. Indeed, when this is done
the anti-ferromagnetic ordering in these 1D arrays is much weaker. Two examples are
shown in Fig. \ref{fig:hilgquad}b,c.
In Fig. \ref{fig:hilgquad}b the half-fluxons all have the same orientation.
We believe that this is because they align with a small residual field.
Fig. \ref{fig:hilgquad}c shows a more random alignment of the isolated half-fluxons.

\begin{figure}
\includegraphics[width=3.5in]{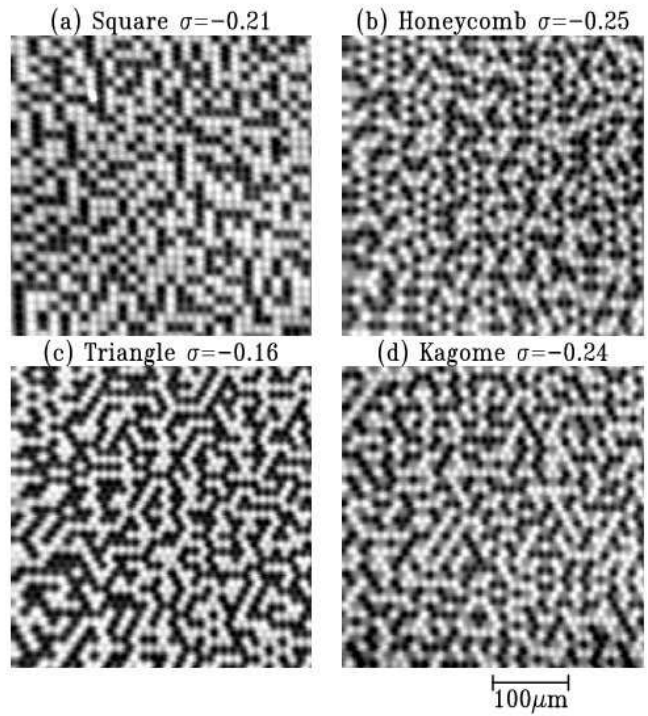}
\vspace{0.1in}
\caption{SQUID microscopy images of 4 electrically disconnected arrays of $\pi$-rings,
in the geometries illustrated in
Fig. \ref{fig:ec1sems}, with values for the ring parameters given in Table I.
These images were taken at 4.2K with a 4$\mu$m diameter pickup loop after cooling in
nominally zero field at 1-10mK/sec.
The spin up fractions $x_+$ and full scale variations in the
scanned SQUID sensor flux ($\Delta\Phi_s$) were
$x_+$=0.56, $\Delta\Phi_s=0.12\Phi_0$ for the square lattice;
$x_+$=0.51, $\Delta\Phi_s=0.11\Phi_0$ for the honeycomb lattice;
$x_+$=0.50, $\Delta\Phi_s=0.10\Phi_0$ for the triangular lattice;
and $x_+$=0.54, $\Delta\Phi_s=0.12\Phi_0$ for the kagome lattice respectively.}
\label{fig:ec1imgs}
\end{figure}

In order to test for magnetic field coupling between the 1D half-fluxon chains,
double facetted junctions (Fig. \ref{fig:hilgfacets}c) were also fabricated and
tested. It appeared that also in this case the phase coupling was stronger than the
field coupling: Fig. \ref{fig:hilgquad}d shows a section of a 40$\mu$m facet double
junction that shows in-phase alignment between the two anti-ferromagnetically ordered
1D chains. Because this arrangement places the positive half-fluxons in the lower
left chain closest to the negative half-fluxons in the upper right chain, this is
the lowest energy arrangement. However, in sections of the 1D chains
which show defects, as in the center of the upper right chains in Fig. \ref{fig:hilgquad}e
and Fig. \ref{fig:hilgquad}f, the interchain alignment goes from in-phase
to out-of-phase when the interchain ordering has a defect, but one chain does not
develop a second defect to align the interchain spins.
Therefore it appears that the
energy cost to create a defect is larger than the energy gain from making neighboring
chains in-phase.

\begin{figure}
\includegraphics[width=3.5in]{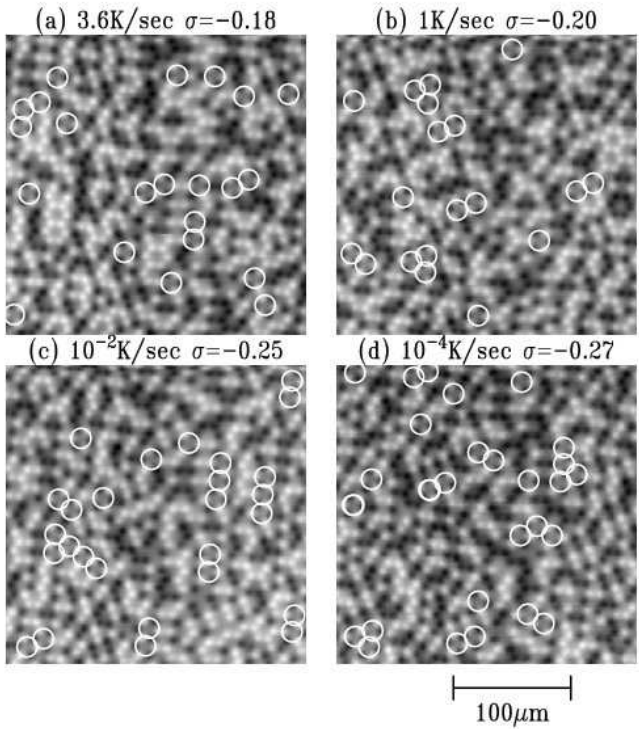}
\vspace{0.1in}
\caption{SQUID microscopy images of a honeycomb array
in the geometry illustrated in
Fig. \ref{fig:ec1sems}, cooled in nominally zero field through the Nb
superconducting transition temperature at different cooling rates.
Each panel is labelled with the cooling rate and experimentally determined
bond order $\sigma$. The white circles superimposed on the images label
6-ring loops in the honeycomb lattice in which the rings are perfectly
anti-ferromagnetically ordered.}
\label{fig:ec1vr}
\end{figure}

When the facetted junctions are cooled in an externally applied magnetic field, one
spin direction becomes energetically favored over the other, but there is a
competition between this energy and the anti-ferromagnetic coupling energy during
the cooling process. An example is shown in Figure \ref{fig:hilgfacs}. More detailed
results and modelling of this facetted junction will be described in the next section.

\begin{figure}
\includegraphics[width=2in]{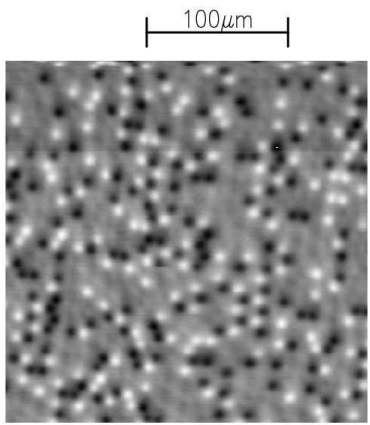}
\vspace{0.1in}
\caption{Difference image obtained from subtracting the image of Fig. \ref{fig:ec1vr}c
from Fig. \ref{fig:ec1vr}b, to determine which rings flipped sign after
successive cooldowns.}
\label{fig:ec1flip}
\end{figure}

In the electrically disconnected 2D lattices of Fig. \ref{fig:ec1sems},
the square and honeycomb arrays are
geometrically unfrustrated, as their magnetic moments can be arranged so that all
nearest neighbors have opposite spins. In contrast, the triangle
and kagom{\'e} lattices are geometrically frustrated, since it is impossible for
all of the rings to have all nearest neighbors anti-ferromagnetically
aligned. Fig. \ref{fig:ec1imgs} shows examples of scanning SQUID microscope images of
the arrays of Fig. \ref{fig:ec1sems} after cooling in nominally zero field.
Although regions of perfect anti-ferromagnetic ordering are seen in the
unfrustrated arrays (Fig. \ref{fig:ec1imgs}a,b),
anti-ferromagnetic ordering beyond a few lattice distances was never observed.
Nevertheless, antiferromagnetic correlations were seen in all the 2D $\pi$-ring arrays.
A measure of the short range antiferromagnetic correlations is the bond order
\begin{equation}
\sigma=1-\frac{x_{AF}}{2x_+x_-},
\label{eq:sigmadef}
\end{equation}
where $x_{AF}$ is the fraction of the nearest neighbor pairs with opposite supercurrent circulation,
and $x_+$($x_-$) is the fraction of rings which have up (down) moments. Perfect antiferromagnetic
correlation would correspond to $\sigma$ = -1. The minimum possible bond order at zero applied field
for the frustrated
triangular and kagom{\'e} lattices is $\sigma$ = -1/3.\cite{davidovic1997}
The images in Fig. \ref{fig:ec1imgs}
are labelled with values for the bond orders.

It is to be expected that the anti-ferromagnetic ordering of the 2D arrays
should improve if they are cooled more slowly through the Nb superconducting
transition. Figure \ref{fig:ec1vr} shows SQUID microscope images of the
same region of the honeycomb 2D lattice of
Fig. \ref{fig:ec1sems}, after cooling at various rates. The individual
panels are labelled with the cooling rates and final state bond orders.
The anti-ferromagnetic ordering increases with slower cooling rates.
One question that can be asked is: Do particular regions of the 2D array order
more strongly than others? The white circles in Fig. \ref{fig:ec1vr} outline
the 6-member rings in the honeycomb arrays in which all neighbors are
anti-ferromagnetically aligned. This provides a convenient way of visualizing
regions of local order. It appears that there are no correlations between
the positions of the ordered 6-member rings from cooldown to cooldown, and we
conclude that the ordered regions are randomly distributed in space.

In the $0$-ring experiments of Davidovic et al. \cite{davidovic1996,davidovic1997}
repeated cooling resulted in a particular ring often being
in the same final state (spin-up or spin-down). This is presumably the result of
the rings having slightly different effective areas, and therefore cooling in
slightly different effective fields, since $0$-rings must be cooled in finite
fields for there to be degenerate states. The individual rings in our 2D $\pi$-ring
arrays appear to have random final states. Figure \ref{fig:ec1flip} shows a
difference image between Fig. \ref{fig:ec1vr}b and Fig. \ref{fig:ec1vr}c,
which were taken after successive cooldowns. In this image, the rings which
did not switch sign from one cooldown to the next are not visible, the rings
which switched from down to up appear black, while the rings which switched
from up to down appear white. Of the roughly 663 rings in this field of view,
152 switched from down to up, and 162 switched from up to down. This is consistent
with random switching during cooldown.

\begin{figure}
\includegraphics[width=3.5in]{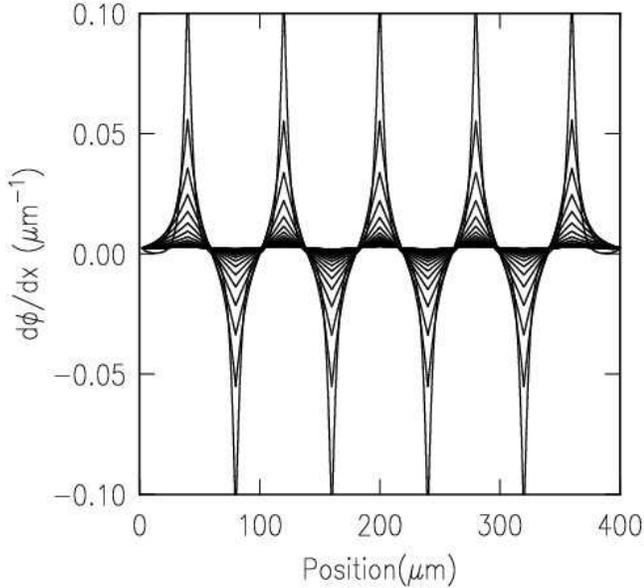}
\vspace{0.1in}
\caption{Modelling of the cooldown of the facetted junction of Fig. \ref{fig:hilgfacs}.
The junction has 10 faces, each 40 microns long, with alternating $0$- and
$\pi$-intrinsic phase shifts.}
\label{fig:oshilg}
\end{figure}

\section{Modelling}

For the current purposes we treat the facetted ramp edge junction as
a linear junction with alternating regions of $0$- and $\pi$- intrinsic phase shifts $\theta(x)$
extending in the $x$-direction, with the junction normal in the $z$ direction, and the
junction width $w$ in the $y$ direction small compared with the
Josephson penetration depth $\lambda_J=\sqrt{\hbar/2 e \mu_0 d j_c}$,
where $d$ is the spacing between the superconducting faces making up the junction,
and $j_c$ is the Josephson critical current per area of the junction.
The quantum mechanical phase drop $\phi(x)$ across the junction is the
solution of the Sine-Gordon equation:
\begin{equation}
\frac{\partial^{2}\phi}{\partial x^{2}} =
\frac{1}{\lambda_{J}^{2}}\sin{( \phi(x)+\theta(x) )}.
\label{eq:sine-gordon}
\end{equation}
Analytical solutions to this equation are available,\cite{xu1995,susanto2003,kogan2000}
but here we solve
Eq. \ref{eq:sine-gordon} numerically.\cite{scalgbj,goldobin2002,goldobin2003,zenchuk2004,goldobin2004c}
Defining a dimensionless coupling parameter
\begin{equation}
\alpha = \frac{\lambda_{J}^{2}}{(\Delta x)^{2}},
\end{equation}
the differential equation Eq. \ref{eq:sine-gordon} turns into a
difference equation on a grid
of size $\Delta x$:
\begin{equation}
{\phi_{n+1}-2\phi_{n}+\phi_{n-1}}
= \frac{1}{\alpha}\sin{(\phi_{n}+\theta_{n})}.
\label{eq:numerical_SG}
\end{equation}
Taking the net current across the junction equal to zero,
using a reduced externally applied magnetic field
$h_e = 2ed\lambda_{J}H_e/\hbar$, and using
the relation between the gradient of the phase and the field $H$ in the
junction:
\begin{equation}
H(x) = \frac{\hbar}{2ed} \frac{\partial \phi}{\partial x},
\end{equation}
we find boundary conditions that are described as difference equations,
where $n_{j}$ is the total number of junctions:
\begin{equation}
\phi_{n_{j}}-\phi_{n_{j}-1}  = \phi_{2}- \phi_{1} = \frac{h_{e}}{\sqrt{\alpha}}.
\end{equation}
These coupled difference equations are solved using a relaxation
method to find the solution $\phi(x)$, iterating to convergence.
As the junction cools through the
superconducting transition temperature T$_c$, the supercurrent density
$j_c$ increases from zero, so that $\lambda_J$ decreases from infinity.
To model the cooling process, we first
solve Eq. \ref{eq:numerical_SG} for $\lambda_J$ much larger than the
facet length $L_f$, then decrease $\lambda_J$ by a small amount, take the
previous solution as the starting point for the next solution, and repeat
until $\lambda_J << L_f$. An example is shown in Fig. \ref{fig:oshilg}.
Here the externally applied flux was set at $\Phi_e=H_edL=\Phi_0/2$, the
total junction length $L$=400$\mu$m, $\Delta x$=1$\mu$m, with 10 facets each
of length $L_f$=40$\mu$m, an initial $\lambda_J=100\mu m$, which was decreased
to 4$\mu$m in 24 equal steps. The field threading the junction is given by
$H(x)=(\Phi_0/2\pi d)d\phi/dx$. The final solution strongly favors anti-ferromagnetic
arrangement of the half-fluxon ``spins".

\begin{figure}
\includegraphics[width=3.5in]{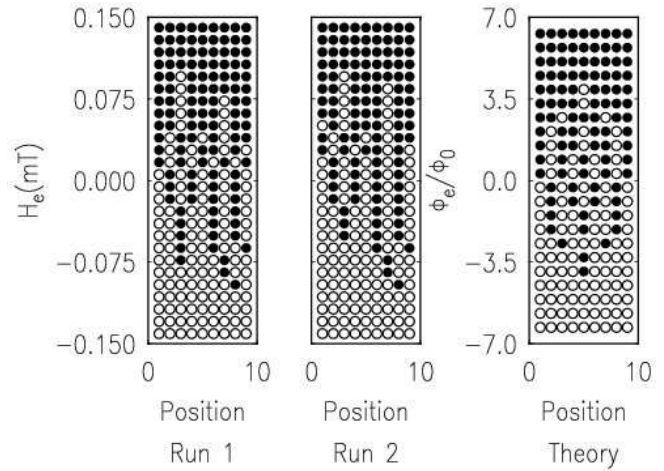}
\vspace{0.1in}
\caption{Results from cooldown of the facetted junction of Fig. \ref{fig:hilgfacs} in
several externally applied magnetic fields. Half-fluxons with
magnetic fields in one direction are represented by open symbols, those with the opposite
sense with closed symbols. The left two panels are experimental results from two successive runs;
the right panel represents modelling as described in the text.}
\label{fig:hilgspi2}
\end{figure}

We can understand why the $0-\pi$ facetted junctions cool into perfect anti-ferromagnetic order,
while the isolated linear arrays of $\pi$-rings do not, by considering the energetics of the cooling process.
Written as a difference equation, the free energy for a particular state of the facetted junction
becomes:
\begin{equation}
F_{V} = \frac{\hbar j_{c} {\rm{w}} \Delta x}{2e}\sum_{1}^{n_{j}-1}
\left( 1-cos(\phi_{n}+\theta_{n})+\frac{\alpha}{2}
(\phi_{n+1}-\phi_{n})^{2} \right)
\end{equation}
Numerical solution of the Sine-Gordon equation as described above
for the facetted junction of Fig. \ref{fig:hilgfacs} shows
that the free energy/facet, when the junction is in perfect anti-ferromagnetic ordering,
is $\sim$ -5$\times$10$^5$K/$\lambda_J$($\mu$m).
The energy cost to form a defect, by flipping one spin, is $\sim$
7.2$\times$10$^5$e$^{-45/\lambda_J (\mu m)}K$.
This implies that when the free energy/facet is comparable to k$_B$T, the energy cost to
form a defect is $\approx$7.2$\times$10$^5$K : it is energetically favorable to form perfect
anti-ferromagnetic ordering in linear $0-\pi$ facetted junctions cooled in zero field.

Figure \ref{fig:hilgspi2} compares the results from repeated cooling of the 10-facet zigzag
junction under various magnetic fields with modelling using the numerical solution of
the sine-Gordon equation outlined above. Note that there are some disorder effects
in the cooling process, as evidenced by the slight differences between the two experimental runs.
The qualitative features of the data are reproduced
by the modelling, although the experimental results are not as symmetric with respect to
inversion in position, or with respect to field reversal, as predicted.
One possible source of the observed asymmetry might be field gradients. However, putting a linear
field gradient into the model did not improve the fit with experiment. Another source of asymmetry
might come from the asymmetry of the junction and lead geometry.

\begin{figure}
\includegraphics[width=3.5in]{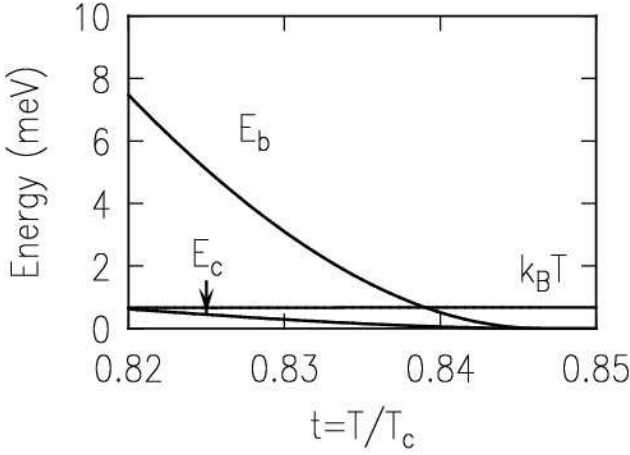}
\vspace{0.1in}
\caption{Calculated energies for the honeycomb ring arrays of Fig. \ref{fig:ec1sems} as a
function of the reduced temperature. $E_b$ is the barrier to thermally activated
flipping of the sign of the spontaneous magnetization, $E_{c}$ is the
strength of the spin-spin coupling energy, and $k_B T$ is the thermal energy.}
\label{fig:hilgengs}
\end{figure}

The cooling process for isolated $\pi$-rings is different than for
electrically connected $\pi$-rings.
For simplicity
we model our isolated $\pi$-rings considering a single inductance $L$ and Josephson
junction critical current $I_c$. The details of the formulas will be different for a two-junction
ring,\cite{smildeprb} but we do not expect the physics to be qualitatively different.
The free energy U of such a $\pi$-ring can be written as
\begin{equation}
U=\frac{\Phi_o^2}{2L}\left\{ \frac{(\Phi-\Phi_e)^2}{\Phi_0^2}
-\frac{LI_c}{\pi\Phi_0}\cos\left(\frac{2\pi\Phi}{\Phi_0}+\pi\right)\right\}
\label{eq:ringenergy}
\end{equation}
where $\Phi$ is the magnetic flux included in the ring, and $\Phi_e$ is the flux externally applied
to the ring.  At temperatures important in the cooling process, close to the Nb superconducting
transition temperature T$_c$, field screening effects can be neglected, since the magnetic
penetration lengths are larger than the size of the sample. Under these conditions, we
estimate the ring self-inductances (for 2.7$\mu$m junction width rings spaced by  11.5$\mu$m) to
be $\approx$3.9pH, and the mutual inductance M between nearest neighbor
rings to be $\approx$0.025pH.
For a single junction $\pi$-ring spontaneous supercurrents flow when $\beta =
2\pi L I_c/\Phi_0$ is just greater than 1. The ring ordering process will occur in the
regime $\beta-1 << 1$. In this limit, and in the limit $\Phi_a/\Phi_0 << 1$, where
$\Phi_a$ is the magnetic flux applied per ring,
the potential barrier to flipping the sign of the ring supercurrents in the $i^{th}$ ring
can be approximated by
\begin{equation}
E_{bi} =3(1-1/\beta_i^\star)^2 I_c \Phi_o/4\pi,
\label{eq:barrier_energy}
\end{equation}
with $\beta_i^\star = \beta + 4\pi\Phi_a\sigma_i/\Phi_0$, where $\sigma_i = \pm1$
is the spin of the $i^{th}$ ring, and
\begin{equation}
\Phi_a = \Phi_e +\sum_j \frac{M_{ij} \Phi_m}{L} \sigma_j,
\label{eq:applied_flux}
\end{equation}
where $\Phi_e$ is the externally applied flux and
$M_{i,j}$ is the mutual inductance between the $i^{th}$ and $j^{th}$ rings.
In the same limits,
the value of the spontaneous flux $\Phi_m$ generated by the $i^{th}$ ring is given by
\begin{equation}
\frac{\Phi_{m,i}}{\Phi_0} = \sqrt{\frac{3}{2\pi^2}(1-1/\beta_i^\star)}.
\label{eq:phim}
\end{equation}
We take the junction critical current
I$_c$(0)=0.55mA for these rings,
and the self-inductance of the rings is assumed to be
temperature independent. A measure of the spin-spin coupling energy $E_c$ is
the difference between the energy barrier for all nearest neighbor spins
up, minus that energy barrier if one of the neighbor spins is down.

Fig. \ref{fig:hilgengs} plots the calculated barrier to thermally activated flipping $E_b$,
nearest neighbor coupling energy $E_c$, and the thermal energy $k_B T$ as a function of
temperature for the honeycomb ring array of Fig. \ref{fig:ec1imgs}.
Both E$_b$ and E$_c$ increase as the temperature is lowered
until the flipping freezes out at a temperature $T_f$ such that
$E_b/k_BT_f \approx -\ln (\gamma(k_B\nu_0 T_f)^{-1}dT/dt)$,
where $\gamma=dE_b(T_f)/dT$.\cite{cosmoprl}
For the rings of Fig. \ref{fig:ec1imgs}, we estimate
this occurs at $(T_{max}-T_f)/T_c$ = 3$\times$10$^{-2}$, with $T_{max}/T_c=1-1/\beta_0$ and
$\beta_0=2\pi L I_c(0)/\Phi_0$.
At this temperature $E_{c}/k_BT_f \approx 1.2$.
Therefore antiferromagnetic coupling is energetically favored in isolated $\pi$-rings, but not nearly
as strongly as in facetted junctions.

\begin{figure}
\includegraphics[width=3.5in]{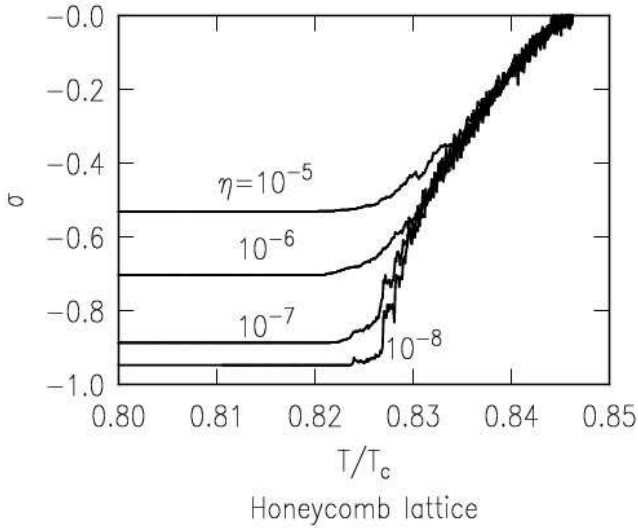}
\vspace{0.1in}
\caption{Bond order as a function of reduced temperature $T/T_c$ calculated
using Monte-Carlo techniques as described in the text, for various cooling
rates $\eta$.}
\label{fig:hilgsvt}
\end{figure}

As the temperature is lowered through $T_c$, the junction critical currents increase
until two distinct circulating states become allowed for $\beta>1$. Thermally activated switching
between these states has a transition rate
\begin{equation}
r \approx \frac{\beta_c}{2\pi\tau}e^{-E_{bi}/k_B T},
\label{eq:rate}
\end{equation}
where $\beta_c=2\pi I_c R^2C/\Phi_0$ is the junction hysteresis parameter, and $\tau \equiv \Phi_0/2\pi I_c R$
is the junction characteristic time.\cite{tesche1981}
We take the junction capacitance $C=5.4\times 10^{-13}F$, $I_c(0)=0.55mA$, $I_c(T)=I_c(0)(1-T/T_c)$,
and $R$=1$\Omega$
for the 2D arrays presented in this paper.
We have taken the limits where the d.c.
shielding currents are much smaller than the critical currents, the spin-flip transition times are much shorter
than the time between spin-flips, and $\beta_c << 1$. We can isolate the temperature dependence of the
prefactor of Eq. \ref{eq:rate} by writing $\beta_c/(2\pi\tau)=4\pi^2I_c^2(0)^2R^3C(1-T/T_c)^2/\Phi_0^2$.
It is of interest to note that the power dissipated per ring near the ``freezout" temperature
is of order $k_B T_c/\tau \sim 10^{-11} W$.

To model the cooling process in our 2D ring arrays we use a Metropolis Monte Carlo\cite{metropolis}
simulation: A 30x30 element array is set up with the same geometry
as the experimental arrays. Boundary effects are minimized by using periodic boundary conditions:
rings at one edge of the array are treated as if their nearest neighbors are the corresponding
rings at the opposite edge. The results of our simulations do not depend sensitively on whether
periodic or non-periodic boundary conditions are used.
For each iteration cycle, the probability of a spin flip $P_i$,
\begin{equation}
P_i=(1-T/T_c)^2e^{-E_{bi}/k_b T},
\label{eq:flipprob}
\end{equation}
is calculated for each ring. A
random number between $0$ and $1$ is generated. If this number is less than $P_i$, the spin
of the ring is flipped. The process is repeated throughout the array, and
the temperature is gradually reduced until no more spin flips
occur. Figure \ref{fig:hilgsvt} shows results from a Monte-Carlo simulation of
the honeycomb lattice of Fig. \ref{fig:ec1sems}, using the ring parameters of
Table I, plotting the bond order $\sigma$
as a function of the reduced temperature $T/T_c$, for various values of $\eta$,
the change in $T/T_c$ per iteration cycle.
Note that in this case the maximum
temperature $T_{max}$ for which spontaneous magnetization is observed is
$T_{max}/T_c=1-1/\beta_0=0.846$. As the temperature is reduced the spin flipping probabilities
decrease, and anti-ferromagnetic ordering gradually occurs.
Furthermore note that the freezing temperature is well below the temperature at which
$E_b \sim k_b T$.
The effective cooling rate of the
simulation is given by $dT/dt = \eta T_c\beta_c(0)/2\pi\tau(0) = 2.2 \times 10^{12} \eta$ K-sec$^{-1}$.
Therefore the slowest cooling rate simulated is about $1 \times 10^{4}$K/sec,
much faster than the experimental cooling rates. Although it would in principle
be possible to use sufficient computer time to match the modelled cooling rates to
experiment, the modelling indicates
stronger anti-ferromagnetic ordering than is observed, even at these very fast cooling rates.

\begin{figure}
\includegraphics[width=3.5in]{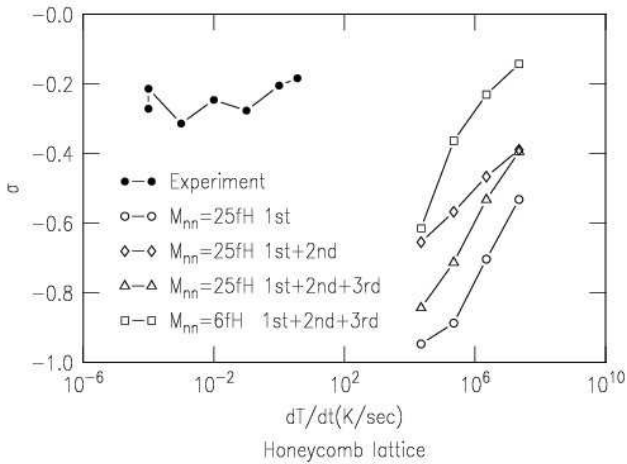}
\vspace{0.1in}
\caption{Low temperature limit of the bond order, as determined experimentally
with SQUID microscopy imaging (solid symbols) for the honeycomb lattice
of Fig. \ref{fig:ec1imgs}, and as calculated for the same lattices
using Monte-Carlo techniques as described in the text (open symbols), for two values of the
mutual inductance between nearest neighbor rings.
The calculations
labelled 1st include only nearest neighbor spin-spin coupling, those labelled 1st+2nd also
include second nearest neighbors, and those labelled 1st+2nd+3rd also include
third nearest neighbors.}
\label{fig:hilgsvr}
\end{figure}

Figure \ref{fig:hilgsvr} shows experimental results for the final bond order for
the honeycomb array of Fig. \ref{fig:ec1imgs}, for various cooling rates. Experimentally
the effect of cooling rate on final bond-order is weak.
Also shown in this Figure are the results
of our Monte-Carlo simulations for the honeycomb array. The points labelled
$M_{nn}=25fH$
use all of the parameters in the simulation as calculated
following Table I.
The points labelled $M_{nn}=6fH$ used a reduced value of the ring-ring
mutual inductance. Also shown are curves comparing the predictions if only
the nearest neighbor spin-spin interactions are included, with results
if 2nd and 3rd nearest neighbors are also included.
The mutual inductance between
a ring and its further neighbors (e.g. 2nd nearest and 3rd nearest) is taken to
be scaled by the cube of the ratio of the relevant center-to-center distances.
Qualitatively, the inclusion
of next nearest neighbors reduces the tendency towards anti-ferromagnetic ordering.
However, it appears that the inclusion of 3rd nearest neighbors does not
change the results much further. Similarly, reducing the strength of the spin-spin coupling
by a factor of 4 also reduces the tendency to
order. However, in all cases it appears that if the simulation were to be extended to
sufficiently slow cooling rates to match experiment, long-range ordering
would result.

\begin{figure}
\includegraphics[width=3.5in]{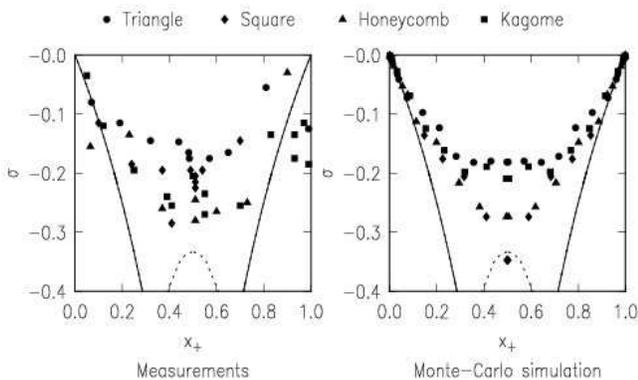}
\vspace{0.1in}
\caption{(a) Measured dependence of the bond order $\sigma$ upon fraction of
spin-up rings $x_+$, for the four different types of arrays shown in Fig. \ref{fig:ec1imgs},
obtained by cooling in various fields. The solid line is the theoretical maximum
negative bond order for a geometrically unfrustrated array; the dashed curve is
that for a frustrated array. Some of the experimental points at very low and high $x_+$ fall below the
theoretical maximum negative bond order because of edge effects due to the finite array sizes
imaged.
(b) Results of a Monte-Carlo simulation of the cooling
process for these arrays, as described in the text. }
\label{fig:hilgrel2}
\end{figure}

This failure of our arrays to order implies that there is a source of disorder that
we haven't taken into account. There are three additional sources of disorder that we have
considered. The first is a distribution in Nb critical temperatures, and the
second is a distribution in junction critical currents. In each case, our simulations
indicate that an unrealistically large distribution width is required to significantly
effect the predicted final $\sigma$ value vs. cooling rate curves. In all the cases
we have explored numerically: reducing $E_b$, reducing $E_c$, increasing the width of the T$_c$
distribution, or increasing the width of the $I_c$ distribution, the $\sigma$ vs. cooling
rate curves can be shifted vertically in Fig. \ref{fig:hilgsvr}, but the slope of these
plots does not change qualitatively. A final factor that might inhibit ordering of our
arrays is if the cooling rate is not uniform. However,
inspection of Fig. \ref{fig:hilgsvt} indicates that fast jumps in temperature of order
0.01$T_c \approx 100mK$ would be required to significantly affect the final bond-orders.
It seems unlikely that our temperature sweeps are that non-uniform.
It is therefore difficult to see how the experimental
data can be fit using the present simulations, and some factor that we haven't considered
correctly is causing the failure of our arrays to achieve long range order.

Nevertheless, the experimentally determined values for the bond-order $\sigma$ can
be modelled fairly well, if one uses the cooling rate as a fitting parameter. The left-hand
panel of Fig. \ref{fig:hilgrel2}a summarizes cooling experiments for all of the 2D arrays of
Fig. \ref{fig:ec1imgs}. These data were taken after cooling the arrays at rates between
1mK/sec and 10mK/sec.  Although long-range anti-ferromagnetic order was not observed
in these arrays, they did show strong anti-ferromagnetic correlations, as evidenced by
the large negative bond orders.  Note that there are no qualitative differences
between the geometrically frustrated and non-frustrated lattices
in their tendency to anti-ferromagnetically order.
For comparison, the $0$-rings of Ref. \onlinecite{davidovic1996,davidovic1997}
never showed
bond-orders more negative than $\sigma = -0.2$, whereas our $\pi$-ring arrays attained
bond orders as negative as $\sigma=-0.3$.
The right hand panel shows the results of Monte-Carlo simulations for
these arrays, using the ring parameters in Table I, except for $M_{nn}=6fH$ instead
of $M_{nn}=25fH$, and with a cooling rate $1 \times 10^6 K/sec$. In
both panels the ideal curve for an unfrustrated lattice (solid line) and a frustrated
lattice (dashed line) are also indicated. The experimental results can be qualitatively
modelled assuming a spin-spin coupling constant about 4 times smaller than calculated,
and a cooling rate $10^{8}$ times faster than the experiment.

\section{Conclusions}

We have shown that it is possible to use large arrays of photolithographically
patterned $\pi$-rings as a model spin system. Half-fluxon Josephson vortices in
electrically connected 1D arrays, in the form
of zigzag $0-\pi$ facetted junctions, order strongly anti-ferromagnetically through
the superconducting order parameter phase. Electrically isolated 1D and 2D arrays order much
less strongly. The 2D $\pi$-ring arrays show stronger anti-ferromagnetic correlations
than reported previously for $0$-ring arrays, but not long-range
order beyond a few lattice constants. We can understand the anti-ferromagnetic phase
coupling of the 1D zigzag junctions by solving the Sine-Gordon equations. Although some
features of the ordering in the 2D arrays can be understood using Monte-Carlo simulations,
unrealistic parameters must be used, and we do not understand why these arrays do not
show long range order.

\section{Acknowledgements}

We would like to thank D.H.A. Blank, R.H. Koch,
K.A. Moler, G. Rijnders, and H. Rogalla for useful discussions.
We also acknowledge the Dutch Foundations FOM and NWO for supporting this research.



\end{document}